\documentclass[dvips]{arxstspdf}
\usepackage{graphics,flushend}
\volume{25}
\issue{1}
\pubyear{2010}
\firstpage{1}
\lastpage{21}
\doi{10.1214/09-STS313}

\begin{document}
\begin{frontmatter}

\title{Matching Methods for Causal Inference: A~Review and a Look Forward}
\runtitle{Matching Methods}

\begin{aug}
\author[a]{\fnms{Elizabeth A.} \snm{Stuart}\ead[label=e1]{estuart@jhsph.edu}\corref{}}
\runauthor{E. A. Stuart}

\affiliation{Johns Hopkins Bloomberg School of Public Health}

\address[a]{Elizabeth A. Stuart is Assistant Professor, Departments of Mental
Health and Biostatistics, Johns Hopkins Bloomberg School of Public Health,
Baltimore, Maryland 21205, USA \printead{e1}.}

\end{aug}

%
\begin{abstract}
When estimating causal effects using observational data, it is
desirable to replicate a randomized experiment as closely as
possible by obtaining treated and control groups with similar
covariate distributions. This goal can often be achieved by
choosing well-matched samples of the original treated and control
groups, thereby reducing bias due to the covariates. Since the
1970s, work on matching methods has examined how to best choose
treated and control subjects for comparison. Matching methods are
gaining popularity in fields such as economics, epidemiology, medicine
and political science. However, until now the literature and related advice
has been scattered across disciplines. Researchers who are interested
in using matching methods---or developing methods related to matching---do
not have a single place to turn to learn about past and current research.
This paper provides a structure for thinking about matching methods and
guidance on their use, coalescing the existing
research (both old and new) and providing
a summary of where the literature on matching methods is now and where
it should be headed.
\end{abstract}

%

\end{frontmatter}
%

\section{Introduction}
\label{sec1}
One of the key benefits of randomized experiments for estimating causal
effects is that the treated and control groups
are guaranteed to be only randomly different from one another on all
background covariates, both observed and unobserved.
Work on matching methods has examined how to
replicate this as much as possible for observed covariates with
observational (nonrandomized) data.
Since early\break work in matching, which began in the 1940s, the methods
have increased in both complexity and
use. However, while the field is expanding, there has been no single
source of information for
researchers interested in an overview of the methods and techniques
available, nor a summary of advice for applied
researchers interested in implementing these methods. In contrast, the
research and resources have been
scattered across disciplines such as statistics (\citeauthor{Rosenbaum02},
\citeyear{Rosenbaum02}; \citeauthor{Rubin06},
\citeyear{Rubin06}),
 epidemiology (\citeauthor{Brookhart06}, \citeyear{Brookhart06}),
  sociology  (Morgan and\break Harding, \citeyear{MorHar06}),
economics (\citeauthor{Imbens04}, \citeyear{Imbens04}) and political science
(\citeauthor{Ho07}, \citeyear{Ho07}).
This paper coalesces the diverse literature on
matching methods, bringing together the original work on matching
methods---of which many current researchers are not aware---and tying
together ideas across disciplines.
In addition to providing guidance on the use of matching methods, the
paper provides a view of where research on matching methods should be headed.

We define ``matching'' broadly to be any method that aims to equate (or
``balance'') the distribution of
covariates in the treated and control groups. This may involve $1\dvtx
1$
matching, weighting or subclassification.
The use of matching methods is in the broader context of the careful
design of nonexperimental studies
(\citeauthor{Rosenbaum99}, \citeyear{Rosenbaum99}, \citeyear{Rosenbaum02}; \citeauthor{Rubin07},\break \citeyear{Rubin07}). While extensive
time and effort is put into the careful design of randomized
experiments, relatively little effort is put into
the corresponding ``design'' of nonexperimental studies. In fact,
precisely because nonexperimental
studies do not have the benefit of randomization, they require even
more careful design.
In this spirit of design, we can think of any study aiming to estimate
the effect of some intervention as having two key stages: (1) design,
and (2) outcome analysis.
Stage (1) uses only background information on the individuals in the
study, designing the nonexperimental study as would be a randomized experiment,
without access to the outcome values. Matching methods are a key tool
for stage~(1). Only after stage (1) is finished does stage (2) begin,
comparing the outcomes of the treated and control individuals. While
matching is generally used
to estimate causal effects, it is also sometimes used for noncausal
questions, for example, to investigate racial disparities (Schneider, Zaslavsky and
Epstein, \citeyear{SchZasEps04}).

Alternatives to matching methods include adjusting for background
variables in a regression
model, instrumental variables, structural equation modeling or
selection models. Matching methods have a few key advantages over those
other approaches.
First, matching methods should not be seen in conflict with regression
adjustment and, in fact, the two methods are complementary and best
used in combination.
Second, matching methods highlight areas of the covariate
distribution where there is not sufficient overlap between the
treatment and control groups, such that the
resulting treatment effect estimates would rely heavily on
extrapolation. Selection models and regression models have been shown
to perform poorly in situations
where there is insufficient overlap, but their standard diagnostics do
not involve checking this overlap
(\citeauthor{DehWah99}, \citeyear{DehWah99}, \citeyear{DehWah02}; \citeauthor{GlaLevMye03}, \citeyear{GlaLevMye03}).\break
Matching methods
in part serve to make researchers aware of the quality of resulting
inferences. Third, matching methods have straightforward diagnostics by
which their performance can be assessed.

The paper proceeds as follows. The remainder of Section \ref{sec1} provides an
introduction to matching methods and the scenarios considered, including
some of the history and theory underlying
matching methods. Sections \ref{sec2}--\ref{sec5} provide details on each of the steps
involved in implementing matching: defining a distance measure,
doing the matching, diagnosing the matching, and then estimating the
treatment effect after matching. The paper concludes with
suggestions for future research and practical guidance in Section \ref{sec6}.

\subsection{Two Settings}
\label{sec1.1}

Matching methods are commonly used in two types of settings. The first
is one in which the outcome values are not yet available and
matching is used to select subjects for follow-up (e.g., \citeauthor{Reinisch95}, \citeyear{Reinisch95}; \citeauthor{StuIal09}, \citeyear{StuIal09}). It is particularly relevant for
studies with cost considerations
that prohibit the collection of outcome data for the full control
group. This was the setting
for most of the original work in matching methods, particularly the
theoretical developments, which compared
the benefits of selecting matched versus random samples of the control group
(\citeauthor{AltRub70}, \citeyear{AltRub70}; \citeauthor{Rubin73a}, \citeyear{Rubin73a}, \citeyear{Rubin73b}).
The second setting is one in which all of the outcome data is already
available, and the goal
of the matching is to reduce bias in the estimation of the treatment effect.

A common feature of matching methods, which is automatic in the first
setting but not the second, is that the outcome values
are not used in the matching process. Even if the outcome values are
available at the time of the matching, the
outcome values should not be used in the matching process. This precludes the selection of a matched sample that leads to a desired
result, or even the appearance of doing so (\citeauthor{Rubin07}, \citeyear{Rubin07}). The
matching can thus be done multiple
times and the matched samples with the best balance---the most similar
treated and control groups---are chosen as the
final matched samples; this is similar to the design of a randomized
experiment where a particular randomization may be rejected if it
yields poor covariate balance (\citeauthor{HilRubTho99}, \citeyear{HilRubTho99}; \citeauthor{Greevy04},
\citeyear{Greevy04}).\looseness=1

This paper focuses on settings with a treatment defined at some
particular point in time, covariates measured at (or relevant to) some period
of time before the treatment, and outcomes
measured after the treatment. It does not consider more complex
longitudinal settings where individuals may go in and out of the
treatment group, or where
treatment assignment date is undefined for the control group.
Methods such as marginal structural models (\citeauthor{RobHerBru00}, \citeyear{RobHerBru00}) or
balanced risk set matching (\citeauthor{LiProRos01}, \citeyear{LiProRos01}) are useful in those settings.

\subsection{Notation and Background: Estimating Causal~Effects}

As first formalized in \citet{Rubin74}, the estimation of causal
effects, whether from a randomized experiment or a nonexperimental study,
is inherently a comparison of potential
outcomes. In particular, the causal effect for individual $i$ is the
comparison of individual
$i$'s outcome if individual $i$ receives the treatment (the potential outcome
under treatment), $Y_i(1)$, and individual $i$'s outcome if individual
$i$ receives the control
(the potential outcome under control), $Y_i(0)$. For simplicity, we use
the term ``individual'' to refer to the units
that receive the treatment of interest, but the formulation would stay
the same if the units were schools or communities. The ``fundamental
problem of
causal inference'' (\citeauthor{Holland86}, \citeyear{Holland86}) is that, for each individual, we
can observe only one of these potential outcomes,
because each unit (each individual at a particular point in time) will
receive either treatment or control, not both.
The estimation of causal effects can thus be thought of as a missing
data problem (\citeauthor{Rubin76c}, \citeyear{Rubin76c}), where we are interested in predicting
the unobserved potential outcomes.

For efficient causal inference and good estimation of the unobserved
potential outcomes,
we would like to compare treated and control groups that are as similar
as possible.
If the groups are very different, the prediction
of $Y(1)$ for the control group will be made using information from
individuals who
look very different from themselves, and likewise for the prediction of
$Y(0)$ for the treated group.
A number of authors, including \citet{CocRub73}, \citeauthor{Rubin73a} (\citeyear{Rubin73a}, \citeyear{Rubin73b}),
\citet{Rubin79},\break \citet{Heckman98b}, Rubin and\break Thomas,
(\citeyear{RubTho00}) and \citet{Rubin01}, have shown that
methods such as
linear regression adjustment can actually increase bias in
the estimated treatment effect when the true relationship between
the covariate and outcome is even moderately nonlinear, especially when
there are large differences in the means and variances
of the covariates in the treated and control groups.

Randomized experiments use a known randomized assignment mechanism to
ensure ``balance'' of
the covariates between the treated and control\break groups: The groups will
be only randomly different from
one another on all covariates, observed and unobserved.
In nonexperimental studies,
we must posit an assignment mechanism, which determines which
individuals receive treatment and which receive
control. A key assumption in nonexperimental studies is that of a
strongly ignorable treatment assignment (\citeauthor{RosRub83a}, \citeyear{RosRub83a}) which
implies that (1)~treatment assignment ($T$) is independent of the
potential outcomes ($Y(0), Y(1)$) given
the covariates ($X$): $ T \bot(Y(0),Y(1)) | X$, and
(2) there is a positive probability of receiving each
treatment for all values of $X$: $ 0 < P(T=1|X) < 1$ for all $X$. The
first component of the definition of strong ignorability is sometimes
termed ``ignorable,'' ``no hidden bias'' or ``unconfounded.'' Weaker versions of the ignorability assumption are sufficient for some
quantities of interest, as discussed further in \citet{Imbens04}.
This assumption is often more reasonable than it may sound at first
since matching on
or controlling for the observed covariates
also matches on or controls for the unobserved covariates, in so much
as they are correlated with those that are
observed. Thus, the only unobserved covariates of concern are those
unrelated to the observed covariates. Analyses can be done to assess
sensitivity of the results to the existence of an
unobserved confounder related to both treatment assignment and the
outcome (see Section~\ref{unobsvar}). \citet{HelRosSma09} also discuss
how matching can make effect estimates less sensitive to an unobserved
confounder, using
a concept called ``design sensitivity.'' An additional assumption
is the Stable Unit Treatment Value\break \mbox{Assumption} (SUTVA; \citeauthor{Rubin80}, \citeyear{Rubin80}), which states that the outcomes of one individual are not
affected by treatment assignment of any other
individuals. While not always plausible---for example, in school
settings where treatment and control children may interact, leading to
``spillover'' effects---the plausibility of SUTVA
can often be improved by design, such as by reducing interactions
between the treated and control groups. Recent work has also begun
thinking about how to relax this assumption in analyses (\citeauthor{HonRau06}, \citeyear{HonRau06};
\citeauthor{Sobel06}, \citeyear{Sobel06};\break
\citeauthor{HudHal08}, \citeyear{HudHal08}).

To formalize, using notation similar to that in \citet{Rubin76a}, we
consider two populations, $P_t$ and $P_c$, where
the subscript $t$ refers to a group exposed to the treatment and $c$
refers to a group
exposed to the control. Covariate data on $p$ pre-treatment
covariates is available on random samples of sizes $N_t$ and $N_c$
from $P_t$ and $P_c$. The
means and variance covariance matrix of the $p$
covariates in group $i$ are given by $\mu_i$ and $\Sigma_i$,
respectively ($i=t,c$). For
individual $j$, the $p$ covariates are denoted by $X_j$, treatment assignment by $T_j$ ($T_j=0$ or 1), and the observed outcome by $Y_j$. Without
loss of generality, we assume $N_t < N_c$.

To define the treatment effect, let $E(Y(1)|X)=R_1(X)$ and
$E(Y(0)|X)=R_0(X)$. In the matching context effects are usually defined
as the difference in potential outcomes, $\tau(x)=R_1(x)-R_0(x)$,\break
although other quantities, such as odds ratios, are also sometimes of interest.
It is often assumed that the response
surfaces, $R_0(x)$ and $R_1(x)$, are parallel, so that \mbox{$\tau(x)=\tau$}
for all $x$.
If the response surfaces are not parallel (i.e., the effect varies), an
average effect over some population is generally estimated. Variation
in effects is
particularly relevant when the estimands of interest are not difference
in means, but rather odds ratios or relative risks, for which the
conditional and marginal effects are not necessarily equal
(\citeauthor{Austin07}, \citeyear{Austin07}; \citeauthor{Lunt09}, \citeyear{Lunt09}). The most common
estimands in nonexperimental studies are the ``average effect of the
treatment on the treated'' (ATT), which is the effect for those in the
treatment group, and the ``average treatment effect'' (ATE),
which is the effect on all individuals (treatment and control). See
\citet{Imbens04}, \citet{Kurth06} and \citet{ImaKinStu07} for further discussion
of these distinctions.
The choice between these estimands will likely involve both substantive
reasons and data availability, as further discussed in Section~\ref{guidance}.

\subsection{History and Theoretical Development of Matching Methods}
\label{history}

Matching methods have been in use since the first half of the 20th
Century (e.g., \citeauthor{Greenwood45}, \citeyear{Greenwood45}; \citeauthor{Chapin47}, \citeyear{Chapin47}), however, a
theoretical basis for these methods was not developed until the
1970s. This development began with papers by \citet{CocRub73} and
\citeauthor{Rubin73a} (\citeyear{Rubin73a}, \citeyear{Rubin73b}) for situations with one covariate and an
implicit focus on estimating the ATT.
\citet{AltRub70} provide an early and excellent discussion
of some practical issues associated with matching: how large
the control ``reservoir'' should be to get good matches, how to define
the quality of matches,
how to define a ``close-enough'' match. Many of the issues identified
in that work are topics of continuing debate and discussion.
The early papers showed that when estimating the ATT,
better matching scenarios include situations with many more control
than treated individuals,
small initial bias between the groups, and smaller variance in the
treatment group than the control group.

Dealing with multiple covariates was a challenge due to both
computational and data problems. With more than just a few
covariates, it becomes very difficult to find matches with close or
exact values of all covariates. For example, \citet{Chapin47} finds
that with initial pools of 671 treated and 523 controls there are only
23 pairs that match exactly on six
categorical covariates. An important advance was made in
1983 with the introduction of the propensity score, defined as the
probability of receiving the treatment given
the observed covariates\break (\citeauthor{RosRub83a}, \citeyear{RosRub83a}). The propensity\break score
facilitates the construction of matched sets with similar distributions of
the covariates, without requiring close or exact matches on all of the
individual variables.

In a series of papers in the 1990s, \citeauthor{RubTho92a} (\citeyear{RubTho92a}, \citeyear{RubTho92b}, \citeyear{RubTho96}) provided
a theoretical basis for multivariate settings with affinely invariant
matching methods and ellipsoidally symmetric covariate distributions
(such as the normal or $t$-distribution), again focusing on estimating
the ATT.
Affinely invariant matching methods, such as propensity score or
Mahalanobis metric matching, are those that yield the same
matches following an affine (linear) transformation of the data. Matching
in this general setting is shown to be Equal Percent Bias Reducing
(EPBR; \citeauthor{Rubin76a}, \citeyear{Rubin76a}).
\citet{RubStu06} later showed that the EPBR feature also holds under
much more general settings, in which the covariate
distributions are discriminant mixtures of ellipsoidally symmetric
distributions. EPBR methods reduce bias in all covariate directions
(i.e., makes the covariate means closer) by the same amount,
ensuring that if close matches are obtained in some direction
(such as the propensity score), then the matching is also reducing bias
in all other directions. The matching thus cannot be increasing bias in an
outcome that is a linear combination of the covariates.
In addition, matching yields the same
percent bias reduction in bias for any linear function of $X$ if and
only if the matching is EPBR.

\citet{RubTho92b} and \citet{RubTho96} obtain
analytic approximations for the reduction in bias on an arbitrary
linear combination of the covariates (e.g., the outcome) that can be
obtained when matching on the true or estimated discriminant (or
propensity score)
with normally distributed covariates.
In fact, the approximations hold remarkably well
even when the distributional assumptions are not satisfied (\citeauthor{RubTho96}, \citeyear{RubTho96}).
The approximations in \citet{RubTho96} can be used to determine in
advance the bias reduction that will be possible from matching, based on
the covariate distributions in the treated and control groups, the size
of the initial difference in the covariates between the groups, the
original sample sizes, the number of matches desired and the
correlation between the covariates and the outcome. Unfortunately these
approximations are rarely used in practice, despite their ability to
help researchers quickly assess whether
their data will be useful for estimating the causal effect of interest.

\subsection{Steps in Implementing Matching Methods}
\label{overview}

Matching methods have four key steps, with the first three representing
the ``design'' and the fourth the ``analysis'': %
\begin{enumerate}
\item Defining ``closeness'': the distance measure used to determine
whether an individual is a good match for another.
\item Implementing a matching method, given that measure of closeness.
\item Assessing the quality of the resulting matched samples, and
perhaps iterating with steps 1 and 2 until
well-matched samples result.
\item Analysis of the outcome and estimation of the treatment effect,
given the matching done in step 3.
\end{enumerate}

The next four sections go through these steps
one at a time, providing an overview of approaches and advice on the
most appropriate methods.

\section{Defining Closeness}\label{sec2}

There are two main aspects to determining the measure of distance (or
``closeness'') to use in matching.
The first involves which covariates to include, and the second involves
combining those covariates into one distance measure.

\subsection{Variables to Include}
\label{variables}
The key concept in determining which covariates to include in the
matching process is that of strong ignorability.
As discussed above, matching
methods, and in fact most nonexperimental study methods, rely on
ignorability, which assumes that there are no unobserved differences
between the treatment and control groups, conditional on the observed
covariates.
To satisfy the assumption of ignorable treatment assignment, it is
important to include in the matching procedure
all variables known to be related to both treatment assignment and
the outcome (\citeauthor{RubTho96}, \citeyear{RubTho96};
Heckman, Ichimura and Todd, \citeyear{Heckman98b};
\citeauthor{GlaLevMye03}, \citeyear{GlaLevMye03};\break
\citeauthor{HilReiZan04}, \citeyear{HilReiZan04}).
Generally poor performance is found of methods that use
a relatively small set of ``predictors of convenience,'' such as
demographics only (\citeauthor{ShaClaSte08}, \citeyear{ShaClaSte08}).
When matching using propensity scores, detailed below, there is little
cost to including variables that are actually unassociated with
treatment assignment, as they will be of little influence in the
propensity score model. Including variables that are actually
unassociated with the outcome can yield slight increases
in variance. However, excluding a potentially important confounder can
be very costly in terms of increased bias. Researchers should thus be
liberal in terms of including
variables that may be associated with treatment assignment and/or
the outcomes. Some examples of matching have 50 or even 100 covariates
included in the procedure (e.g., \citeauthor{Rubin01}, \citeyear{Rubin01}). However, in small
samples it may not be possible to include a very large set of
variables. In that case priority should be given to variables believed
to be related to the outcome, as there is a higher cost in terms of
increased variance of including variables unrelated to the outcome but
highly related to treatment assignment (\citeauthor{Brookhart06}, \citeyear{Brookhart06}).
Another effective strategy is to include a small set of covariates
known to be related to the outcomes of interest, do the matching, and
then check the balance on all
of the available covariates, including any additional variables that
remain particularly unbalanced after the matching.
To avoid allegations of variable selection based on estimated effects,
it is best if the variable selection process is done without
using the observed outcomes, and instead is based on previous research
and scientific understanding (\citeauthor{Rubin01}, \citeyear{Rubin01}).

One type of variable that should not be included in the matching
process is any variable that may have been affected by the treatment of
interest\break (\citeauthor{Rosenbaum84b}, \citeyear{Rosenbaum84b}; \citeauthor{FraRub02},
\citeyear{FraRub02};\break \citeauthor{Greenland03}, \citeyear{Greenland03}).
This is especially important when the covariates, treatment indicator
and outcomes are all collected at the same
point in time. If it is deemed to be critical to control for
a variable potentially affected by treatment assignment, it is better to exclude that variable in the matching
procedure and include it in the analysis model for the outcome
(as in \citeauthor{Reinisch95}, \citeyear{Reinisch95}).\footnote{The method is misstated in the
footnote in Table 1 of that paper. In fact, the potential confounding variables
were not used in the matching procedure, but were utilized in the
outcome analysis (D.~B. Rubin, personal communication).}

Another challenge that potentially arises is when variables are fully
(or nearly fully) predictive of treatment assignment. Excluding
such a variable\break  should be done only with great
care, with the belief that the problematic variable is completely
unassociated with the outcomes of interest and that the ignorability
assumption will still hold. More commonly, such a variable
indicates a fundamental problem in estimating the effect of interest,
whereby it may not be possible to separate out the effect of the
treatment of interest from this problematic variable using
the data at hand. For example, if all adolescent heavy drug users are
also heavy drinkers, it will be impossible to separate out the effect
of heavy drug use from the effect of heavy drinking.

\subsection{Distance Measures}
\label{distances}
The next step is to define the ``distance'': a measure of the similarity
between two individuals.
There are four primary ways to define the distance $D_{ij}$ between
individuals $i$ and $j$ for matching, all of which are affinely invariant:
\begin{enumerate}
\item Exact:
\[
D_{ij} =
\cases{
 0, & if $X_i=X_j$,\vspace*{2pt} \cr
\infty,& if $ X_i \neq X_j$.}
\]
\item Mahalanobis:
\[
D_{ij} = (X_i - X_j)' \Sigma^{-1} (X_i - X_j).
\]
If interest is in the ATT, $\Sigma$ is the variance covariance matrix
of $X$ in the full control group; if interest is in the
ATE, then $\Sigma$ is the variance covariance matrix of $X$ in the
pooled treatment and full control groups. If $X$ contains categorical variables,
they should be converted to a series of binary indicators, although the
distance works best with continuous variables.
\item Propensity score:
\[
D_{ij} = |e_i - e_j|,
\]
where $e_k$ is the propensity score for individual $k$, defined in
detail below.
\item Linear propensity score:
\[
D_{ij} = |\operatorname{logit}(e_i) - \operatorname{logit}(e_j)|.
\]
\citet{RosRub85a}, \citet{RubTho96} and \citet{Rubin01} have found that
matching on the linear propensity score can be particularly effective
in terms of reducing bias.
\end{enumerate}
Below we use ``propensity score'' to refer to either the propensity
score itself or the linear version.

Although exact matching is in many ways the ideal (\citeauthor{ImaKinStu07}, \citeyear{ImaKinStu07}),
the primary difficulty with the exact and Mahalanobis distance measures
is that neither works very well when
$X$ is high dimensional. Requiring exact matches often leads to many
individuals not being matched, which can result
in larger bias than if the matches are inexact but more individuals
remain in the analysis\break (\citeauthor{RosRub85a}, \citeyear{RosRub85a}). A recent
advance, coarsened exact matching (CEM), can be used to do exact
matching on broader ranges of the variables; for example, using income
categories rather than a continuous measure (\citeauthor{IacKinPor09}, \citeyear{IacKinPor09}).
The Mahalanobis distance can work quite well when there are relatively
few covariates
(fewer than 8; \citeauthor{Rubin79}, \citeyear{Rubin79}; \citeauthor{Zhao04}, \citeyear{Zhao04}), but it does not perform as
well when the covariates are not normally distributed or there are many
covariates\break (\citeauthor{GuRos93}, \citeyear{GuRos93}).
This is likely because Mahalanobis
metric matching essentially regards all interactions among the elements
of $X$ as equally important; with more covariates, Mahalanobis matching
thus tries to match
more and more of these multi-way interactions.

A major advance was made in 1983 with the introduction of propensity
scores (\citeauthor{RosRub83a}, \citeyear{RosRub83a}). Propensity
scores summarize all of the covariates into one scalar: the probability
of being treated.
The propensity score for individual $i$
is defined as the probability of receiving the treatment
given the observed covariates: $e_i(X_i)=P(T_i=1|X_i)$.
There are two key properties of propensity scores. The first is that
propensity scores are balancing scores: At each value of the propensity
score, the distribution of the covariates $X$ defining
the propensity score is the same in the
treated and control groups. Thus, grouping individuals with similar
propensity scores replicates a mini-randomized experiment, at least
with respect to the observed
covariates. Second, if treatment assignment is ignorable
given the covariates, then treatment assignment is also ignorable given
the propensity score. This justifies matching based on the propensity
score rather than on the full
multivariate set of
covariates. Thus, when treatment assignment is ignorable, the
difference in means in the outcome between
treated and control individuals with a particular propensity score
value is an unbiased estimate of the treatment effect at that propensity
score value.
While most of the propensity score results are in the context of finite
samples and the settings considered by\break \citeauthor{RubTho92a} (\citeyear{RubTho92a},
\citeyear{RubTho96}), Abadie and Imbens (\citeyear{AbaImb07}) discuss the asymptotic properties of propensity score matching.

The distance measures described above can also be combined, for
example, doing exact matching on key covariates such as race or gender
followed by propensity score matching
within those groups. When exact matching on even a few variables is not
possible because of sample size limitations, methods that yield
``fine balance'' (e.g., the same proportion of African American males in
the matched treated and control groups) may be a good alternative (\citeauthor{RosRosSil07}, \citeyear{RosRosSil07}).
If the key covariates of interest are continuous, Mahalanobis matching
within propensity score calipers\break (\citeauthor{RubTho00}, \citeyear{RubTho00}) defines the distance
between individuals $i$ and $j$ as
\[
D_{ij} =
\cases{
 (Z_i - Z_j)' \Sigma^{-1} (Z_i - Z_j),\vspace*{2pt} \cr
  \quad\hspace*{16pt} \mbox{if }|\operatorname{logit}(e_i)-\operatorname{logit}(e_j)| \leq c,\vspace*{2pt} \cr
\infty,\quad  \mbox{if }|\operatorname{logit}(e_i)-\operatorname{logit}(e_j)| >
c,}
\]
where $c$ is the caliper, $Z$ is the set of ``key covariates,'' and
$\Sigma$ is the variance covariance matrix of $Z$.
This will yield matches that are relatively well matched on the
propensity score and particularly well matched on~$Z$. $Z$ often
consists of
pre-treatment measures of the outcome, such as baseline test scores in
educational evaluations.
\citet{RosRub85a} discuss the choice of caliper size, generalizing
results from Table~2.3.1 of \citet{CocRub73}.
When the variance of the linear propensity score in the treatment group
is twice as large as that in the control group,
a caliper of $0.2$ standard deviations removes $98\%$ of the bias in a normally
distributed covariate. If the variance in the treatment group is much
larger than that in the control group, smaller calipers are necessary.
\citet{RosRub85a}
generally suggest a caliper of 0.25 standard deviations of the linear
propensity score.

A more recently developed distance measure is the ``prognosis score'' (\citeauthor{Hansen08}, \citeyear{Hansen08}). Prognosis scores are essentially the predicted outcome
each individual would have under the control condition. The benefit of
prognosis scores is that they take into account the relationship
between the covariates
and the outcome; the drawback is that it requires a model for that
relationship. Since it thus does not have the clear separation of the
design and analysis
stages that we advocate here, we focus instead on other approaches, but
it is a potentially important advance in the matching literature.

\subsubsection{Propensity score estimation and model specification}
\label{pscoreest}

In practice, the true propensity scores are rarely known outside of
randomized experiments and thus must be estimated. Any model relating a
binary variable to a set
of predictors can be used. The most common for propensity score
estimation is logistic regression, although nonparametric methods such
as boosted CART
and generalized boosted models (gbm) often show very good performance
(\citeauthor{McCRidMor04}, \citeyear{McCRidMor04};\break \citeauthor{Setoguchi08},
\citeyear{Setoguchi08}; \citeauthor{LeeLesStu09}, \citeyear{LeeLesStu09}).

The model diagnostics when estimating propensity scores are not the
standard model diagnostics
for logistic regression or CART. With propensity score estimation,
concern is not with the parameter estimates
of the model, but rather with the resulting balance of the covariates
(\citeauthor{AugSch00}, \citeyear{AugSch00}). Because of this, standard concerns
about\break
collinearity do not apply. Similarly, since they do not use covariate
balance as a criterion, model fit statistics identifying classification
ability (such as the $c$-statistic) or stepwise selection models are not
helpful for variable selection (\citeauthor{Rubin04}, \citeyear{Rubin04};
\citeauthor{Brookhart06}, \citeyear{Brookhart06};
\citeauthor{Setoguchi08}, \citeyear{Setoguchi08}).
One strategy that is helpful is to examine the balance of covariates
(including those not originally included in the propensity score
model), their squares and interactions in the matched samples.
If imbalance is found on particular variables or functions of
variables, those terms can be included in a re-estimated propensity
score model, which should improve their balance in the
subsequent matched samples (\citeauthor{RosRub84a}, \citeyear{RosRub84a}; \citeauthor{DehWah02}, \citeyear{DehWah02}).

Research indicates that misestimation of the propensity score (e.g.,
excluding a squared term that is in the true model) is not a large problem,
and that treatment effect estimates are more biased when the outcome
model is misspecified than when
the propensity score model is misspecified (\citeauthor{Drake93},
\citeyear{Drake93};\break
\citeauthor{DehWah99}, \citeyear{DehWah99}, \citeyear{DehWah02};
\citeauthor{Zhao04}, \citeyear{Zhao04}). This may in part
be because the propensity score is used only as a tool to get covariate
balance---the accuracy of the model is less important as long as balance
is obtained. Thus, the exclusion of
a squared term, for example, may have less severe consequences for a
propensity score model than it does for the outcome model, where
interest is in interpreting a particular
regression coefficient (that on the treatment indicator).
However, these evaluations are fairly limited; for example, \citet
{Drake93} considers
only two covariates. Future research should involve more systematic
evaluations of propensity score estimation, perhaps through more
sophisticated simulations
as well as analytic work, and consideration should include how the
propensity scores will be used, for example, in weighting versus
subclassification.

\section{Matching Methods}\label{sec3}

Once a distance measure has been selected, the next step is to use that
distance in doing the matching.
In this section we provide an overview of the spectrum of matching
methods available. The methods primarily vary in terms of
the number of individuals that remain after matching and in the
relative weights that different individuals receive.
One way in which propensity scores are commonly used is as a predictor
in the outcome model, where the set of individual covariates is
replaced by the propensity score
and the outcome models run in the full treated and control groups (\citeauthor{Weitzen04},
\citeyear{Weitzen04}).
Unfortunately the simple use of this method is not an optimal use of
propensity scores, as it does not take advantage of the balancing
property of propensity scores: If there is imbalance on the original
covariates, there will also be imbalance on the propensity score,
resulting in the same degree of model extrapolation as with the full
set of covariates. However, if the model regressing the outcome on the
treatment indicator and the propensity score is correctly
specified or if it includes nonlinear functions of the propensity score
(such as quantiles or splines) and their interaction with the treatment
indicator, then this can be an effective approach, with
links to subclassification (\citeauthor{SchKan08}, \citeyear{SchKan08}). Since this method does not
have the clear ``design'' aspect of matching, we do not discuss it further.

\subsection{Nearest Neighbor Matching}
\label{nearest}

One of the most common, and easiest to implement and understand,
methods is $k\dvtx 1$ nearest
neighbor matching (\citeauthor{Rubin73a}, \citeyear{Rubin73a}). This is generally the most
effective method for settings where the goal is to select
individuals for follow-up. Nearest neighbor matching nearly always
estimates the ATT, as it matches control individuals to the treated
group and discards controls who are not selected as matches.

In its simplest form, $1\dvtx 1$ nearest neighbor matching selects for each
treated individual $i$ the
control individual with the smallest distance from individual~$i$. A~common complaint regarding $1\dvtx 1$ matching
is that it can discard a large number of observations and thus would
apparently lead to reduced power. However, the reduction in power is
often minimal, for two main reasons.
First, in a two-sample comparison of means, the precision is largely
driven by the smaller group size (\citeauthor{Cohen88}, \citeyear{Cohen88}). So if the treatment
group stays the same size, and only the control
group decreases in size, the overall power may not actually be reduced
very much (\citeauthor{Ho07}, \citeyear{Ho07}). Second, the power increases when the groups are
more similar because of the reduced extrapolation and higher precision
that is obtained when comparing groups that are similar versus groups that
are quite different (\citeauthor{SneCoc80}, \citeyear{SneCoc80}). This is also what yields the
increased power of using matched pairs in randomized experiments\break (\citeauthor{WacWei82}, \citeyear{WacWei82}).
\citet{Smith97} provides an illustration where estimates from $1\dvtx 1$
matching have lower standard deviations than estimates from a linear
regression, even though thousands of observations were discarded in the
matching.
An additional concern is that, without any restrictions, $k\dvtx 1$ matching
can lead to some poor matches, if, for example, there are no control
individuals with propensity scores similar to a given treated
individual. One strategy to avoid poor matches is to impose a caliper
and only select a match if it is within the caliper.
This can lead to difficulties in interpreting effects if many treated
individuals do not receive a match, but can help avoid poor matches.
\citet{RosRub85b} discuss those trade-offs.

\subsubsection{Optimal matching}
\label{optimal}

One complication of simple (``greedy'') nearest neighbor matching is
that the order in which the treated subjects are matched may change the
quality of the matches.
Optimal matching avoids this issue by taking into account the overall
set of matches when choosing individual
matches, minimizing a global distance measure (\citeauthor{Rosenbaum02}, \citeyear{Rosenbaum02}).
Generally, greedy matching performs poorly when there is intense
competition for controls, and performs well when there
is little competition\break (\citeauthor{GuRos93}, \ \citeyear{GuRos93}).  \ \citet{GuRos93} find that optimal
matching does not in general perform any better than greedy matching in
terms of creating groups with good balance, but
does do better at reducing the distance within pairs (page 413): ``\ldots optimal
matching picks about the same controls [as greedy matching] but does a
better job of assigning them to treated units.'' Thus, if the goal is
simply to find well-matched groups,
greedy matching may be sufficient. However, if the goal is well-matched
pairs, then optimal matching may be preferable.

\subsubsection{Selecting the number of matches: Ratio\break matching}
\label{ratiofull}

When there are large numbers of control individuals, it is sometimes
possible to get multiple good matches for each treated individual,
called ratio matching (\citeauthor{Smith97}, \citeyear{Smith97}; \citeauthor{RubTho00}, \citeyear{RubTho00}). Selecting the
number of matches involves a bias$\dvtx $variance trade-off.
Selecting multiple controls
for each treated individual will generally increase bias since the 2nd,
3rd and 4th closest matches are, by definition, further away from the
treated individual
than is the 1st closest match. On the other hand, utilizing multiple
matches can decrease variance due to the larger matched sample size.
Approximations
in \citet{RubTho96} can help determine the best ratio. In settings
where the outcome data has yet to be collected and there are cost
constraints, researchers must also balance cost considerations. More
methodological work needs to be done to
more formally quantify the trade-offs involved. In addition, $k\dvtx 1$
matching is not optimal since it does not account for the fact that
some treated individuals may have many close matches while others have
very few. A more advanced form
of ratio matching, variable ratio matching, allows the ratio to vary,
with different treated individuals
receiving differing numbers of matches\break (\citeauthor{MinRos01}, \citeyear{MinRos01}). Variable ratio
matching is related to full matching, described below.

\subsubsection{With or without replacement}
\label{replacement}

Another key issue is whether controls can be used as matches for more
than one treated individual: whether
the matching should be done ``with replacement'' or\break ``without
replacement.'' Matching with replacement can often decrease bias
because controls that look similar to many treated individuals can be
used multiple times.
This is particularly helpful in settings where there are few control
individuals comparable to the treated individuals (e.g., \citeauthor{DehWah99}, \citeyear{DehWah99}). Additionally,
when matching with replacement, the order in which the treated
individuals are matched does not matter. However, inference becomes more
complex when matching with replacement, because the matched controls
are no longer independent---some are in the matched sample more than
once and
this needs to be accounted for in the outcome analysis, for example, by
using frequency weights. When
matching with replacement, it is also possible that the treatment
effect estimate will be based on just a small number of controls; the
number of times each control is matched should be monitored.

\subsection{Subclassification, Full Matching and Weighting}

For settings where the outcome data is already available, one apparent
drawback of $k\dvtx 1$ nearest neighbor matching is that it does not
necessarily use all the data, in that
some control individuals, even some of those with propensity scores in
the range of the treatment groups' scores, are discarded and not
used in the analysis. Weighting, full matching and subclassification
methods instead use all individuals. These methods can be thought of as
giving all individuals
(either implicit or explicit) weights between 0 and 1, in contrast with
nearest neighbor matching, in which individuals essentially receive a
weight of either 0 or 1 (depending on whether or not they are selected
as a match). The three methods discussed here represent a continuum in
terms of the number of groupings formed, with weighting as the limit of
subclassification as the number of
observations and subclasses go to infinity (\citeauthor{Rubin01}, \citeyear{Rubin01}) and full
matching in between.

\subsubsection{Subclassification}
\label{subclassification}

Subclassification forms\break   groups of individuals who are similar, for
example, as defined by quintiles of the propensity
score distribution. It can estimate either the ATE or the ATT, as
discussed further in Section \ref{sec5}. One of the first uses of
subclassification was \citet{Cochran68}, which examined
subclassification on a single covariate (age) in investigating the link
between lung cancer and smoking.
\citet{Cochran68} provides analytic expressions
for the bias reduction possible using subclassification on a
univariate continuous covariate; using just five subclasses removes at
least $90\%$ of the initial
bias due to that covariate. \citet{RosRub85a} extended that to show that
creating five propensity score subclasses
removes at least 90\% of the bias in the estimated treatment effect due to
all of the covariates that went into the propensity score.
Based on those results, the current convention is to use 5--10
subclasses. However, with larger sample sizes more subclasses (e.g.,
10--20) may be feasible and appropriate (\citeauthor{LunDav04}, \citeyear{LunDav04}).
More work needs to be done to help determine the optimal number of
subclasses: enough to get adequate bias reduction but not too many that
the within-subclass effect estimates
become unstable.

\subsubsection{Full matching}

A more sophisticated form of subclassification, full matching, selects
the number of subclasses automatically (\citeauthor{Rosenbaum91b},
\citeyear{Rosenbaum91b};\break
\citeauthor{Hansen04}, \citeyear{Hansen04};
\citeauthor{StuGre08}, \citeyear{StuGre08}). Full matching
creates a series of matched sets,
where each\break matched set contains at least one treated individual and at
least one control individual (and each matched set may have many from
either group).
Like subclassification, full matching can estimate either the ATE or
the ATT. Full matching is optimal in terms of minimizing the average of
the distances between each treated individual
and each control individual within each matched set. \citet{Hansen04}
demonstrates the method in the context of estimating the effect of SAT
coaching. In that example the original treated and control groups had
propensity score differences of 1.1 standard deviations, but the matched
sets from full matching differed by only 0.01 to 0.02 standard
deviations. Full matching may
thus have appeal for researchers who are reluctant to discard some of
the control individuals but who want to obtain optimal balance on the
propensity score. To achieve efficiency gains, \citet{Hansen04} also
introduces restricted ratios of the number of treated individuals to
the number of control individuals in each matched set.

\subsubsection{Weighting adjustments}

Propensity scores\break can also be used directly as inverse weights in
estimates of the ATE, known as inverse probability of treatment
weighting (IPTW; \citeauthor{Czajka92},\break \citeyear{Czajka92};
\citeauthor{RobHerBru00}, \citeyear{RobHerBru00};\break
\citeauthor{LunDav04}, \citeyear{LunDav04}).
Formally, the weight $w_i = \frac{T_i}{\hat{e}_i} + \frac{1-T_i}{1-\hat
{e}_i}$, where $\hat{e}_k$\vspace*{1pt} is the estimated
propensity score for individual $k$. This weighting serves to weight
both the treated and control groups up to the full sample, in the same
way that survey sampling weights weight a sample up to
a population (\citeauthor{HorTho52}, \citeyear{HorTho52}).

An alternative weighting technique, weighting by the odds, can be used
to estimate the ATT\break (\citeauthor{HirImbRid03}, \citeyear{HirImbRid03}). Formally,
$w_i = T_i + (1-T_i)\frac{\hat{e}_i}{1-\hat{e}_i}.$
With this weight, treated individuals receive a weight of 1. Control
individuals are weighted up to the full sample using the $\frac
{1}{1-\hat{e}_i}$ term, and then weighted to the treated
group using the $\hat{e}_i$ term. In this way both groups are weighted
to represent the treatment group.

A third weighting technique, used primarily in economics, is kernel
weighting, which averages over multiple individuals
in the control group for each treated individual, with weights defined
by their distance (\citeauthor{Imbens00}, \citeyear{Imbens00}).
\citet{Heckman97}, \citet{Heckman98a} and  Heckman, Ichimura and Todd (\citeyear{Heckman98b}) describe a local linear
matching estimator that requires specifying a bandwidth parameter.
Generally, larger bandwidths
increase bias but reduce variance by putting weight on individuals that
are further away from the treated individual of interest. A
complication with these methods
is this need to define a bandwidth or smoothing parameter, which does
not generally have an intuitive meaning; \citet{Imbens04} provides
some guidance on that choice.

A potential drawback of the weighting approaches is that, as with
Horvitz--Thompson estimation, the variance
can be very large if the weights are extreme (i.e., if the estimated
propensity scores are close to 0 or~1). If the model is correctly
specified and thus the weights are correct, then the large variance is
appropriate. However, a worry is that some of the extreme weights may
be related more to the estimation procedure
than to the true underlying probabilities. Weight trimming, which sets
weights above some maximum to that maximum, has been proposed as one
solution to this problem (\citeauthor{Potter93}, \citeyear{Potter93}; \citeauthor{SchRotRob99}, \citeyear{SchRotRob99}). However, there
is relatively little guidance regarding the trimming level. Because of
this sensitivity to the size of the weights and potential model
misspecification,
more attention should be paid to the accuracy of propensity score
estimates when the propensity\break scores will be used for weighting vs.
matching\break (\citeauthor{KanSch07}, \citeyear{KanSch07}). Another effective strategy
is doubly-robust methods (\citeauthor{BanRob05},\break \citeyear{BanRob05}), which yield accurate effect
estimates if either the propensity score model or the outcome model are
correctly specified, as discussed
further in Section~\ref{sec5}.

\subsection{Assessing Common Support}
\label{commsupp}

One issue that comes up for all matching methods is that of ``common support.'' To this point, we have assumed that there is substantial overlap of the
propensity score distributions in the two groups, but potentially
density differences. However, in some situations there may not be
complete overlap in the distributions. For example, many of the control
individuals
may be very different from all of the treatment group members, making
them inappropriate as points of comparison when estimating the ATT
(\citeauthor{AusMam06}, \citeyear{AusMam06}). Nearest neighbor\break matching with calipers automatically
only uses individuals in (or close to) the area of common support.
In contrast, the subclassification and weighting methods generally use
all individuals, regardless of the overlap of the distributions.
When using those methods it may be beneficial to explicitly\break restrict
the analysis to those individuals\break in the region of common support
(as in\break \citeauthor{Heckman97}, \citeyear{Heckman97};\break
\citeauthor{DehWah99}, \citeyear{DehWah99}).

Most analyses define common support using the propensity score, discarding
individuals with propensity score values outside the range of the other group.
A~second method involves examining the ``convex hull'' of the
covariates, identifying the multidimensional space that allows
interpolation rather than extrapolation
(\citeauthor{KinZen06}, \citeyear{KinZen06}). While these procedures can help identify who needs to
be discarded, when many subjects are discarded it can help the
interpretation of results if it is possible to define the discard rule
using one or two covariates rather than the propensity score itself.

It is also important to consider the implications of common support for
the estimand of interest.
Examining the common support may indicate that it is not possible to
reliably estimate the ATE. This could happen, for example,
if there are controls outside the range of the treated individuals and
thus no way to estimate $Y(1)$ for the controls without extensive
extrapolation. When estimating the ATT it may
be fine (and in fact beneficial) to discard controls outside the range
of the treated individuals, but discarding treated individuals may
change the group for which the results apply
(\citeauthor{Crump09}, \citeyear{Crump09}).

\section{Diagnosing Matches}
\label{sec4}
Perhaps the most important step in using matching methods is to
diagnose the quality of the resulting matched samples.
All matching should be followed by an assessment of the covariate balance
in the matched groups, where balance is defined as the similarity of
the empirical distributions of the full set of covariates in the
matched treated and control groups. In other words,
we would like the treatment to be unrelated to the covariates, such
that $\tilde{p}(X|T=1)=\tilde{p}(X|T=0)$, where $\tilde{p}$ denotes the
empirical distribution.
A matching method that results in highly imbalanced samples should be
rejected, and alternative methods should
be attempted until a well-balanced sample is attained. In some
situations the diagnostics may indicate that the treated and control
groups are too far apart to
provide reliable estimates without heroic modeling assumptions (e.g., \citeauthor{Rubin01}, \citeyear{Rubin01};
\citeauthor{AgoDyn03}, \citeyear{AgoDyn03}). In contrast to traditional regression
models, which do not
examine the joint distribution of the predictors (and, in particular,
of treatment assignment and the covariates), matching methods will
make it clear when it is not possible to separate
the effect of the treatment from other differences between the groups.
A well-specified regression model of the outcome with many interactions
would show this imbalance and
may be an effective method for estimating treatment effects (\citeauthor{SchKan08}, \citeyear{SchKan08}), but complex models like that are only rarely used.

When assessing balance we would ideally compare the multidimensional
histograms of the covariates in the matched treated and control groups.
However, multidimensional histograms are very coarse and/or will have
many zero cells. We thus are left examining the balance of
lower-dimensional summaries of that
joint distribution, such as the marginal distributions of each
covariate. Since we are attempting to examine different features of the
multidimensional distribution, though, it is
helpful to do a number of different types of balance checks, to obtain
a more complete picture.

All balance metrics should be calculated in ways similar to how the
outcome analyses will be run, as discussed further in
Section \ref{sec5}. For example, if subclassification was done, the
balance measures should be calculated within each subclass and then
aggregated. If weights will be used in analyses (either as IPTW or
because of variable ratio or full matching), they should also be used
in calculating the balance measures (\citeauthor{Joffe04}, \citeyear{Joffe04}).

\subsection{Numerical Diagnostics}
\label{numbalance}
One of the most common numerical balance diagnostics is the difference
in means of each covariate, divided by the standard
deviation in the full treated group: $\frac{\overline{X}_{t}-\overline
{X}_{c}}{\sigma_{t}}$. This measure, sometimes referred to as the
``standardized bias'' or ``standardized
difference in means,'' is similar to an effect size and is compared
before and after matching\break (\citeauthor{RosRub85a}, \citeyear{RosRub85a}). The same standard
deviation should be used in the standardization before and after
matching. The standardized difference of means should be computed for
each covariate, as well as two-way interactions
and squares. For binary covariates, either this same formula can be
used (treating them as if they were continuous), or a simple difference
in proportions can be calculated (\citeauthor{Austin09}, \citeyear{Austin09}).

\citet{Rubin01} presents three balance measures\break based on the theory in
\citet{RubTho96} that provide a comprehensive view of covariate
balance:
\begin{enumerate}
\item The standardized difference of means of the\break propensity score.
\item The ratio of the variances of the propensity score in the treated
and control groups.
\item For each covariate, the ratio of the variance of the residuals
orthogonal to the propensity score in the treated and control groups.
\end{enumerate}
\citet{Rubin01} illustrates these diagnostics in an example with 146 covariates.
For regression adjustment to be trustworthy, the absolute standardized
differences of means should be less than 0.25 and the variance ratios
should be between 0.5 and 2 (\citeauthor{Rubin01}, \citeyear{Rubin01}).
These guidelines are based both on the assumptions underlying
regression adjustment as well as on results in \citet{Rubin73b} and \citet
{CocRub73}, which used simulations to estimate the bias resulting from
a number of treatment effect estimation procedures when the true relationship
between the covariates and outcome is even moderately nonlinear.

Although common, hypothesis tests and $p$-values that incorporate
information on the sample size (e.g., $t$-tests) should not be used as
measures of\break balance, for two main reasons (\citeauthor{Austin07}, \citeyear{Austin07};\break
\citeauthor{ImaKinStu07}, \citeyear{ImaKinStu07}). First, balance
is inherently an in-sample property, without reference to any broader
population or super-population. Second, hypothesis tests can be
misleading as measures of balance, because they often conflate changes
in balance with changes in statistical power.\break \citet{ImaKinStu07} show
an example where
randomly discarding control individuals seemingly leads to increased
balance, simply because of the reduced power. In particular, hypothesis
tests should not be used as part of a stopping
rule to select a matched sample when those samples have varying sizes
(or effective sample sizes). Some researchers argue that hypothesis
tests are okay for testing balance since
the outcome analysis will also have reduced power for estimating the
treatment effect (Hansen, \citeyear{Hansen08n}), but that argument requires trading off Type I and
Type II errors. The cost of those two types of errors
may differ for balance checking and treatment effect estimation.

\subsection{Graphical Diagnostics}

\begin{figure}[b]

\includegraphics{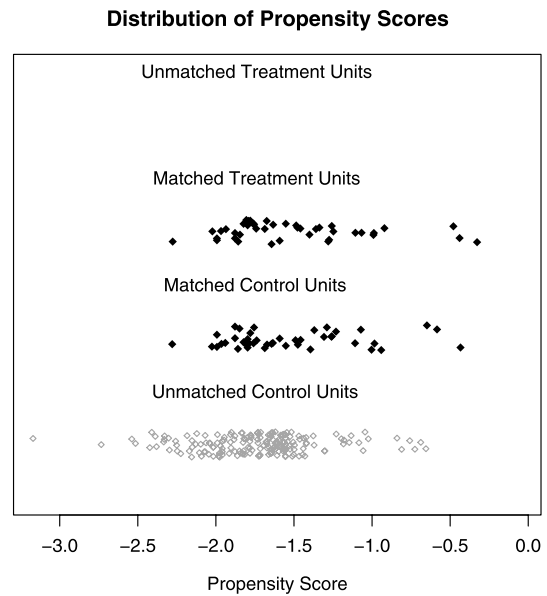}

\caption{Matches chosen using $1\dvtx 1$ nearest neighbor matching on propensity
score. Black dots indicate matched individuals; grey unmatched
individuals. Data from Stuart and Green (\protect\citeyear{StuGre08}).}
\label{figurejitter}
\end{figure}

With many covariates it can be difficult to carefully examine numeric
diagnostics for each; graphical diagnostics can be helpful for getting
a quick assessment of the covariate balance. A first step is to examine
the distribution of the propensity scores
in the original and matched groups; this is also useful for assessing
common support. Figure \ref{figurejitter} shows an example with
adequate overlap of
the propensity scores, with a good control match for each treated
individual. For weighting or subclassification, plots such as this can
show the
dots with their size proportional to their weight.

For continuous covariates, we can also examine quantile--quantile (QQ)
plots, which compare the empirical distributions of each variable in
the treated and control groups (this could
also be done for the variables squared or two-way interactions, getting
at second moments). QQ plots compare the quantiles of a variable in the
treatment group
against the corresponding quantiles in the control group. If the two
groups have identical empirical
distributions, all points would lie on the 45 degree line. For
weighting methods, weighted boxplots can provide similar information
(\citeauthor{Joffe04}, \citeyear{Joffe04}).

\begin{figure}

\includegraphics{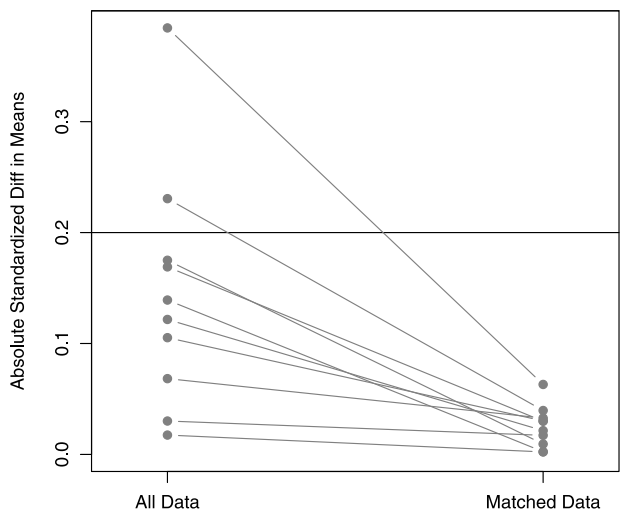}

\caption{Plot of standardized difference of means of 10 covariates
before and after matching. Data from Stuart and Green (\protect\citeyear{StuGre08}).}
\label{figurestdbias}
\end{figure}

Finally, a plot of the standardized differences of means, as in Figure
\ref{figurestdbias}, gives us a quick overview of whether balance has
improved for individual covariates (\citeauthor{RidMcCMor06}, \citeyear{RidMcCMor06}).
In\break this example the standardized difference of means of each covariate
has decreased after matching. In some situations researchers may find
that the standardized difference of means of a few covariates will
increase. This may be particularly true of covariates with small
differences before matching, since they will not factor heavily into
the propensity
score model (since they are not predictive of treatment assignment). In these cases researchers should consider whether the
increase in bias on those covariates is problematic, which it may be
if those covariates are strongly related to the outcome, and modify the
matching accordingly (\citeauthor{Ho07}, \citeyear{Ho07}). One solution for that may be to do
Mahalanobis matching on those covariates within
propensity score calipers.

\section{Analysis of the Outcome}
\label{sec5}

Matching methods are not themselves methods for estimating causal
effects. After the matching has created treated and control groups with
adequate balance
(and the observational study thus\break ``designed''), researchers can move to
the outcome analysis stage. This stage will
generally involve regression adjustments using the matched samples,
with the details of the analysis depending on the structure of the matching.
A key point is that matching methods are not designed to ``compete''
with modeling adjustments such as linear regression, and, in fact,
the two methods have been shown to work best in combination (\citeauthor{Rubin73b}, \citeyear{Rubin73b};
\citeauthor{Carpenter77}, \citeyear{Carpenter77};
\citeauthor{Rubin79}, \citeyear{Rubin79};
\citeauthor{RobRot95}, \citeyear{RobRot95};\break
\citeauthor{Heckman97}, \citeyear{Heckman97};
Rubin and\break Thomas, \citeyear{RubTho00};
\citeauthor{GlaLevMye03}, \citeyear{GlaLevMye03};
\citeauthor{AbaImb06}, \citeyear{AbaImb06}).
This is similar to the idea of ``double robustness,'' and the intuition
is the same as that behind regression adjustment in randomized
experiments, where
the regression adjustment is used to ``clean up'' small residual
covariate imbalance between the groups. Matching methods should also make
the treatment effect estimates less sensitive to particular outcome
model specifications (\citeauthor{Ho07}, \citeyear{Ho07}).


The following sections describe how outcome analyses should proceed
after each of the major types of matching methods described above. When
weighting
methods are used, the weights are used directly in regression models,
for example, using weighted least squares. We focus on parametric
modeling approaches since
those are the most commonly used, however, nonparametric
permutation-based tests,\break such as Fisher's exact test, are also
appropriate, as detailed in \citeauthor{Rosenbaum02} (\citeyear{Rosenbaum02,Rosenbaum10}). The best results
are found when estimating marginal treatment effects, such as
differences in means or differences in proportions. \citet{GreRobPea99}
and \citet{Austin07} discuss some of the challenges in estimating
noncollapsible conditional treatment effects and which matching methods
perform best for those situations.

\subsection{After $k\dvtx 1$ Matching}

When each treated individual has received $k$\break  matches, the outcome
analysis proceeds using the matched samples, as if those samples had
been generated through randomization.
There is debate about whether the analysis needs to account for the
matched pair nature of the data (\citeauthor{Austin07}, \citeyear{Austin07}). However, there are at
least two reasons why it is not necessary to account for the matched
pairs (\citeauthor{SchKan08}, \citeyear{SchKan08}; \citeauthor{Stuart07b}, \citeyear{Stuart07b}). First, conditioning on the variables
that were used in the matching process (such as through a regression
model) is
sufficient. Second, propensity score matching, in fact, does not
guarantee that the individual pairs will be well-matched on the full
set of covariates, only that groups of individuals with similar
propensity scores will have similar covariate distributions. Thus, it
is more common to simply pool all the matches into matched treated and
control groups and run analyses using the groups as a whole, rather
than using the individual matched pairs.
In essence, researchers can do the exact same analysis they would have
done using the original data,
but using the matched data instead (\citeauthor{Ho07}, \citeyear{Ho07}).

Weights need to be incorporated into the analysis for matching with
replacement or variable\break ratio matching (\citeauthor{DehWah99}, \citeyear{DehWah99};\break
\citeauthor{HilReiZan04}, \citeyear{HilReiZan04}).
When matching\break with replacement, control group individuals receive a
frequency weight that reflects the number of times they were selected
as a match. When using variable ratio matching, control group members
receive a weight that is proportional to the number of controls matched
to ``their'' treated
individual. For example, if 1 treated individual was matched to 3
controls, each of those controls receives a weight of $1/3$. If another
treated individual was matched to just 1 control, that control receives
a weight of 1.

\subsection{After Subclassification or Full Matching}

With standard subclassification (e.g., the formation of 5 subclasses),
effects are generally estimated within each subclass and then
aggregated across subclasses (\citeauthor{RosRub84a}, \citeyear{RosRub84a}).
Weighting the subclass estimates by the number of treated individuals
in each subclass estimates the ATT; weighting by the overall number of
individuals in each subclass estimates the ATE. There may be fairly
substantial imbalance remaining in each subclass and, thus,
it is important to do regression adjustment within each subclass, with
the treatment indicator and covariates as predictors (\citeauthor{LunDav04}, \citeyear{LunDav04}).
When the number of subclasses is too large---and the number of
individuals within each subclass too small---to estimate separate
regression models within each subclass,
a joint model can be fit, with subclass and subclass by treatment
indicators (fixed effects). This is especially useful for full matching.
This estimates a separate effect for each subclass, but assumes that
the relationship
between the covariates $X$ and the outcome is constant across
subclasses. Specifically, models such as $Y_{ij} = \beta_{0j} + \beta
_{1j}T_{ij} + {\gamma X_{ij}} + e_{ij}$ are fit,
where $i$ indexes individuals and $j$ indexes subclasses. In this
model, $\beta_{1j}$ is the treatment effect for subclass $j$, and these
effects are aggregated across subclasses to obtain an overall
treatment effect: $\beta= \frac{N_j}{N} \sum_{j=1}^{J} \beta_{1j}$,
where $J$\vspace*{1pt} is the number of subclasses, $N_j$ is the number of
individuals in subclass $j$, and $N$ is the total number of individuals.
(This formula weights subclasses by their total size, and so estimates
the ATE, but could be modified to estimate the ATT.)
This procedure is somewhat more complicated for noncontinuous outcomes
when the estimand of interest, for example, an odds ratio, is noncollapsible.
In that case the outcome proportions in each treatment group should be
aggregated and then combined.

\subsection{Variance Estimation}

One of the most debated topics in the literature on matching is
variance estimation. Researchers disagree on whether uncertainty in the
propensity
score estimation or the matching procedure needs to be taken into
account, and, if so, how. Some researchers (e.g., \citeauthor{Ho07}, \citeyear{Ho07}) adopt
an approach similar to randomized experiments,
where the models are run conditional on the covariates, which are
treated as fixed and exogenous. Uncertainty regarding the matching
process is not taken into account. Other researchers argue that
uncertainty in the
propensity score model needs to be accounted for in any analysis.\break
However, in fact, under fairly general conditions\break (\citeauthor{RubTho96}, \citeyear{RubTho96}; \citeauthor{RubStu06}, \citeyear{RubStu06}), using estimated rather than true
propensity scores leads to an overestimate of variance, implying that
not accounting for the uncertainty in using estimated rather than true
values will be conservative in the sense of yielding confidence
intervals that are wider than necessary. \citet{RobMarNew92} also show
the benefit of using estimated rather than true propensity scores.
Analytic expressions for the bias and variance reduction possible for
these situations are given in \citet{RubTho92b}.
Specifically, \citet{RubTho92b} states that ``\ldots with large pools of
controls, matching using estimated
linear propensity scores results in approximately half the variance for
the difference in the
matched sample means as in corresponding random samples for all
covariates uncorrelated with
the population discriminant.'' This finding has been confirmed in simulations
(\citeauthor{RubTho96}, \citeyear{RubTho96}) and an empirical example (\citeauthor{HilRubTho99}, \citeyear{HilRubTho99}). Thus,
when it is possible to obtain $100\%$ or nearly $100\%$ bias
reduction by matching on true or estimated propensity scores, using the
estimated propensity scores will result in more precise estimates of
the average
treatment effect. The intuition is that the estimated propensity score
accounts for chance imbalances between the groups, in addition to the
systematic differences---a situation where overfitting is good.
When researchers want to account for the uncertainty in the matching, a
bootstrap procedure has been found to outperform other methods (\citeauthor{Lechner02}, \citeyear{Lechner02}; \citeauthor{HilRei06}, \citeyear{HilRei06}).
There are also some empirical formulas for variance estimation for
particular matching scenarios (e.g., \citeauthor{AbaImb06}, \citeyear{AbaImb06}, \citeyear{AbaImb09b};
\citeauthor{SchKan08}, \citeyear{SchKan08}), but this is an area for future research.

\section{Discussion}\label{sec6}

\subsection{Additional Issues}

This section raises additional issues that arise when using any
matching method, and also provides suggestions for future research.

\subsubsection{Missing covariate values}

Most of the literature on matching and propensity scores assume fully
observed covariates, but of course most studies have at least some
missing data.
One possibility is to use generalized boosted models to estimate
propensity scores, as they do not require fully observed covariates.
Another recommended approach is to do a simple single imputation of the
missing covariates and include missing data indicators in the
propensity score model. This essentially
matches based both on the observed values and on the missing data
patterns. Although this is generally not an appropriate strategy for
dealing with missing data (\citeauthor{GreFin95}, \citeyear{GreFin95}), it
is an effective approach in the propensity score context. Although it
cannot balance the missing values themselves, this method will yield
balance on the observed covariates
and the missing data patterns (Rosenbaum and Rubin, \citeyear{RosRub84a}). A more flexible method
is to use multiple imputation to impute the missing covariates, run the
matching and effect estimation separately within each ``complete'' data
set, and then use the multiple imputation combining rules to obtain
final effect estimates (\citeauthor{Rubin87}, \citeyear{Rubin87}; \citeauthor{Song01}, \citeyear{Song01}). \citet{QuLip09}
illustrate this method and show good results for an adaptation that
also includes indicators of missing data patterns in the propensity
score model.

In addition to development and investigation of matching methods that
account for missing data, one particular area needing development is
balance diagnostics for settings with missing covariate values,
including dignostics that allow for nonignorable missing data
mechanisms. \citet{DagRub00} suggests
a few simple diagnostics such as assessing available-case means and
standard deviations of the continuous variables,
and comparing available-case cell proportions for the categorical
variables and missing-data indicators, but diagnostics should be
developed that explicitly consider the interactions
between the missing data and treatment assignment mechanisms.

\subsubsection{Violation of ignorable treatment assignment}
\label{unobsvar}

A critique of any nonexperimental study is that there may be unobserved
variables related to both treatment assignment and the outcome,
violating the assumption of ignorable treatment assignment
and biasing the treatment effect estimates. Since ignorability can
never be directly tested, researchers
have instead developed sensitivity analyses to assess its plausibility,
and how violations of ignorability may affect study conclusions. One
type of plausibility test estimates an effect on a variable that is known
to be unrelated to the treatment, such as a pre-treatment measure of
the outcome variable (as in \citeauthor{Imbens04}, \citeyear{Imbens04}), or the difference in
outcomes between
multiple control groups (as in \citeauthor{Rosenbaum87a}, \citeyear{Rosenbaum87a}). If the test
indicates that the effect is not equal to zero, then the assumption of
ignorable treatment assignment is deemed
to be less plausible.

A second approach is to perform analyses of sensitivity to an
unobserved variable.
Rosenbaum and Rubin (\citeyear{RosRub83b}) extends the ideas of \citet{Cornfield59}, examining how
strong the correlations would have to be between a hypothetical
unobserved covariate and both treatment assignment and the outcome
to make the observed
treatment effect go away. Similarly, bounds can be created for the
treatment effect, given a range of potential correlations of the
unobserved covariate with treatment assignment
and the outcome (\citeauthor{Rosenbaum02}, \citeyear{Rosenbaum02}).
Although sensitivity analysis methods are becoming more and more
developed, they are still used relatively\break infrequently. Newly available
software\break (\citeauthor{McCRidMor04}, \citeyear{McCRidMor04};
\citeauthor{Keele09},\break \citeyear{Keele09}) will hopefully help facilitate
their adoption by more researchers.

\subsubsection{Choosing between methods}

There are a wide variety of matching methods available, and little
guidance to help applied researchers select between them (Section~\ref
{guidance} makes an attempt). The primary advice to this point
has been to select the method that yields the best balance (e.g., Harder, Stuart and Anthony, \citeyear{Harder10};
\citeauthor{Ho07}, \citeyear{Ho07};
\citeauthor{Rubin07}, \citeyear{Rubin07}). But defining the best balance is
complex, as it involves trading off balance on multiple covariates.
Possible ways to choose a method include the following: (1) the method
that yields the smallest standardized difference of means across the
largest number of
covariates, (2) the method that minimizes the standardized difference
of means of a few particularly prognostic covariates, and (3) the
method that results in the fewest number of
``large'' standardized differences of means (greater than 0.25). Another
promising direction is work by \citet{DiaSek05}, which automates the
matching procedure, finding the best matches according to a set of
balance measures. Further research needs to compare the performance of
treatment effect estimates from methods using criteria such as those
in\break \citet{DiaSek05} and  Harder, Stuart and Anthony (\citeyear{Harder10}), to determine what the proper
criteria should be and examine issues such as potential overfitting to
particular measures.

\subsubsection{Multiple treatment doses}
\label{multdoses}
Throughout this\break discussion of matching, it has been assumed that there
are just two groups: treated and control. However, in many
studies there are actually multiple levels of the treatment (e.g.,
doses of a drug).\break
\citet{Rosenbaum02} summarizes two methods for dealing with doses of
treatment. In the first method, the propensity
score is still a scalar function of the covariates (e.g., \citeauthor{JofRos99}, \citeyear{JofRos99}; \citeauthor{LuZanRos01}, \citeyear{LuZanRos01}).
In the second method, each
of the levels of treatment has its own propensity score (e.g.,\break \citeauthor{Rosenbaum87b}, \citeyear{Rosenbaum87b}; \citeauthor{Imbens00}, \citeyear{Imbens00})
 and each
propensity score is used one at a time to estimate the distribution of
responses that would have been observed if all individuals had received
that dose.

Encompassing these two approaches,\break \citet{ImaVan04} generalizes the
propensity score to arbitrary treatment regimes
(including ordinal, categorical and multidimensional).
They provide theorems for the properties of this generalized propensity
score (the propensity function), showing that it has properties
similar to that of the propensity score in that adjusting for the
low-dimensional (not always
scalar, but always low-dimensional)\break
 propensity function balances the
covariates. They advocate subclassification rather than matching,
and provide two examples as well as simulations showing the performance
of adjustment based on the propensity
function. Diagnostics are also complicated in this setting, as it
becomes more difficult to assess the balance of the
resulting samples when there are multiple treatment levels. Future work
is needed to
examine these issues.

\subsection{Guidance for Practice}
\label{guidance}

So what are the take-away points and advice regarding when to use each
of the many methods discussed? While more work is needed to
definitively answer that question, this section attempts to pull
together the current literature to provide advice for researchers
interested in estimating causal effects using matching methods. The
lessons can be summarized as follows:
\begin{longlist}
\item[1.] Think carefully about the set of covariates to include in the
matching procedure, and err on the side of\vadjust{\goodbreak} including more rather than
fewer. Is the ignorability assumption reasonable given
that set of covariates? If not, consider in advance whether there are
other data sets that may be more appropriate, or if there are
sensitivity analyses that can be done to strengthen the inferences.
\item[2.] Estimate the distance measure that will be used in the matching.
Linear propensity scores estimated using logistic regression, or
propensity scores estimated using generalized boosted models or\break boosted
CART, are good choices. If there are a few covariates on which
particularly close balance is desired (e.g., pre-treatment measures of
the outcome), consider using the Mahalanobis distance within\break propensity
score calipers.
\item[3.] Examine the common support and implications for the estimand. If
the ATE is of substantive interest, is there enough overlap of the
treated and control groups' propensity scores to estimate the ATE?
If not, could the ATT be estimated more reliably? If the ATT is of
interest, are there controls across the full range of the treated
group, or will it be difficult to estimate the effect
for some treated individuals?
\item[4.] Implement a matching method.
\begin{itemize}
\item If estimating the ATE, good choices are generally IPTW or full matching.
\item If estimating the ATT and there are many more control than
treated individuals (e.g., more than 3 times as many), $k\dvtx 1$ nearest
neighbor matching without replacement is a good choice for its
simplicity and good performance.
\item If estimating the ATT and there are not (or not many) more
control than treated individuals, appropriate choices are generally
subclassification, full matching and weighting by the odds.
\end{itemize}
\item[5.] Examine the balance on covariates resulting from that matching method.
\begin{itemize}
\item If adequate, move forward with treatment effect estimation, using
regression adjustment on the matched samples.
\item If imbalance on just a few covariates, consider incorporating
exact or Mahalanobis matching on those variables.
\item If imbalance on quite a few covariates, try another matching
method (e.g., move to $k\dvtx 1$ matching with replacement) or consider
changing the estimand or the data.
\end{itemize}
\end{longlist}

Even if for some reason effect estimates will not be obtained using
matching methods, it is worthwhile to go through the steps outlined
here to assess the adequacy of the data for answering the question of interest.
Standard regression diagnostics will not warn researchers when there is
insufficient overlap to reliably estimate causal effects; going through
the process of estimating propensity scores and assessing balance
before and after matching can be invaluable in terms of helping
researchers move forward with causal inference with confidence.

Matching methods are important tools for applied researchers and also
have many open research questions for statistical development. This
paper has provided an overview of the current literature on matching
methods, guidance for practice and a road map for future research. Much
research has been done in the past 30 years
on this topic, however, there are still a number of open areas and
questions to be answered.
We hope that this paper, combining results from a variety of
disciplines, will promote awareness of and interest in matching methods
as an important and interesting area for future research.

\section{Software Appendix}\label{sec7}

In previous years software limitations made it\break  \mbox{difficult} to implement
many of the more advanced matching methods. However, recent advances
have made these methods more and more accessible.\break  This section lists
some of the major matching procedures available. A continuously updated
version\break is  also available at
\href{http://www.biostat.jhsph.edu/\textasciitilde estuart/propensityscoresoftware.html}{http://www.biostat.jhsph.edu/}
\href{http://www.biostat.jhsph.edu/\textasciitilde estuart/propensityscoresoftware.html}{\textasciitilde estuart/propensityscoresoftware.html}.
\begin{longlist}
\item[$\bullet$] Matching software for R
\begin{itemize}
\item[--]\textbf{cem}, \url{http://gking.harvard.edu/cem/}
\item[]
Iacus, S. M., King, G. and Porro, G. (2009). cem: Coarsened exact
matching software. Can also be implemented through MatchIt.

\item[--]\textbf{Matching},
\href{http://sekhon.berkeley.edu/matching}{http://sekhon.berkeley.edu/matching}
\item[]Sekhon, J. S. (in press). Matching: Multivariate and propensity score
matching with balance optimization. Forthcoming, \textit{Journal of
Statistical Software.} Uses automated procedure to select matches,
based on univariate and multivariate balance diagnostics.
Primarily $k\dvtx 1$ matching, allows matching with or without replacement,
caliper, exact. Includes built-in effect and variance estimation
procedures.

\item[--]\textbf{MatchIt}, \url{http://gking.harvard.edu/matchit}
\item[]
Ho, D. E., Imai, K., King, G. and Stuart, E. A. (in press). MatchIt:
Nonparametric preprocessing for parameteric causal inference.
Forthcoming, \textit{Journal of Statistical Software.} Two-step process:
does matching, then user does outcome analysis.
Wide array of estimation procedures and matching methods available:
nearest neighbor, Mahalanobis, caliper, exact, full, optimal,
subclassification. Built-in numeric and graphical diagnostics.

\item[--]\textbf{optmatch},
\href{http://cran.r-project.org/web/packages/optmatch/index.html}{http://cran.r-project.org/web/}\break
\href{http://cran.r-project.org/web/packages/optmatch/index.html}{packages/optmatch/index.html}
\item[]
Hansen, B. B. and Fredrickson, M. (2009). optmatch: Functions for
optimal matching. Variable ratio, optimal and full matching. Can also
be implemented through MatchIt.

\item[--]\textbf{PSAgraphics},
\href{http://cran.r-project.org/web/packages/PSAgraphics/index.html}{http://cran.r-project.org/web/}\break
\href{http://cran.r-project.org/web/packages/PSAgraphics/index.html}{packages/PSAgraphics/index.html}
\item[]
Helmreich, J. E. and Pruzek, R. M. (2009). PSAgraphics: Propensity score
analysis graphics. \textit{Journal of Statistical Software} \textbf{29}. Package
to do graphical diagnostics of propensity score methods.

\item[--]\textbf{rbounds},
\href{http://cran.r-project.org/web/packages/rbounds/index.html}{http://cran.r-project.org/web/packages/}
\href{http://cran.r-project.org/web/packages/rbounds/index.html}{rbounds/index.html}
\item[]
Keele, L. J. (2009). rbounds: An R package for sensitivity analysis with
matched data. Does analysis of sensitivity to assumption of ignorable
treatment assignment.

\item[--]\textbf{twang},
\href{http://cran.r-project.org/web/packages/twang/index.html}{http://cran.r-project.org/web/packages/}\break
\href{http://cran.r-project.org/web/packages/twang/index.html}{twang/index.html}
\item[]
Ridgeway, G., McCaffrey, D. and Morral, A. (2006). twang: Toolkit for
weighting and analysis of\break nonequivalent groups. Functions for
propensity\break score estimating and weighting, nonresponse weighting, and
diagnosis of the weights. Primarily uses generalized boosted regression
to estimate the\break propensity scores.
\end{itemize}

\item[$\bullet$] Matching software for Stata
\begin{itemize}
\item[--]\textbf{cem}, \url{http://gking.harvard.edu/cem/}
\item[]
Iacus, S. M., King, G. and Porro, G. (2009). cem: Coarsened exact
matching software.

\item[--]\textbf{match},
\href{http://www.economics.harvard.edu/faculty/imbens/software\_imbens}{http://www.economics.harvard.edu/faculty/}
\href{http://www.economics.harvard.edu/faculty/imbens/software\_imbens}{imbens/software\_imbens}
\item[] Abadie, A., Drukker, D., Herr, J. L. and Imbens, G.~W. (2004).
Implementing matching estimators for average treatment effects in
Stata. \textit{The Stata Journal} \textbf{4} 290--311. Primarily $k\dvtx 1$ matching
(with replacement). Allows estimation of ATT or ATE, including robust
variance estimators.

\item[--]\textbf{pscore},
\href{http://www.lrz-muenchen.de/\textasciitilde sobecker/pscore.html}{http://www.lrz-muenchen.de/\textasciitilde
sobecker/}\break
\href{http://www.lrz-muenchen.de/\textasciitilde sobecker/pscore.html}{pscore.html}
\item[]Becker, S. and Ichino, A. (2002). Estimation of average treatment
effects based on propensity scores. \textit{The Stata Journal} \textbf{2}
358--377. Does $k\dvtx 1$ nearest neighbor matching, radius (caliper) matching
and subclassification.

\item[--]\textbf{psmatch2},
\href{http://econpapers.repec.org/software/bocbocode/s432001.htm}{http://econpapers.repec.org/software/}
\href{http://econpapers.repec.org/software/bocbocode/s432001.htm}{bocbocode/s432001.htm}
\item[]Leuven, E. and Sianesi, B. (2003). psmatch2. Stata module to perform
full Mahalanobis and propensity score matching, common support
graphing, and covariate imbalance testing. Allows $k\dvtx 1$ matching, kernel
weighting, Mahalanobis matching. Includes built-in diagnostics
and procedures for estimating ATT or ATE.

\item[--]\textbf{Note}: 3 procedures for analysis of sensitivity to the
ignorability assumption are also available:\break rbounds (for continuous
outcomes), mhbounds (for categorical outcomes), and sensatt (to be used after
the pscore procedures).

\item[]\textbf{rbounds},
\href{http://econpapers.repec.org/software/bocbocode/s438301.htm}{http://econpapers.repec.org/software/}\break
\href{http://econpapers.repec.org/software/bocbocode/s438301.htm}{bocbocode/s438301.htm};

\item[]\textbf{mhbounds},
\href{http://ideas.repec.org/p/diw/diwwpp/dp659.html}{http://ideas.repec.org/p/diw/diwwpp/}
\href{http://ideas.repec.org/p/diw/diwwpp/dp659.html}{dp659.html};

\item[]\textbf{sensatt},
\href{http://ideas.repec.org/c/boc/bocode/s456747.html}{http://ideas.repec.org/c/boc/bocode/}\break
\href{http://ideas.repec.org/c/boc/bocode/s456747.html}{s456747.html}.
\end{itemize}

\item[$\bullet$] Matching software for SAS
\begin{itemize}
\item[--] \textbf{SAS usage note}:
\href{http://support.sas.com/kb/30/971.html}{http://support.sas.com/kb/30/}
\href{http://support.sas.com/kb/30/971.html}{971.html}

\item[--] \textbf{Greedy $\bolds{1\dvtx 1}$ matching},
\href{http://www2.sas.com/proceedings/sugi25/25/po/25p225.pdf}{http://www2.sas.com/}\break
\href{http://www2.sas.com/proceedings/sugi25/25/po/25p225.pdf}{proceedings/sugi25/25/po/25p225.pdf}

\item[]
Parsons, L. S. (2005). Using SAS software to perform a case-control
match on propensity score in an observational study. In \textit{SAS SUGI 30},
Paper 225-25.

\item[--] \textbf{gmatch macro},
\href{http://mayoresearch.mayo.edu/mayo/research/biostat/upload/gmatch.sas}{http://mayoresearch.mayo.edu/mayo/}
\href{http://mayoresearch.mayo.edu/mayo/research/biostat/upload/gmatch.sas}{research/biostat/upload/gmatch.sas}
\item[]Kosanke, J. and Bergstralh, E. (2004). gmatch: Match 1 or more
controls to cases using the\break  GREEDY algorithm.

\item[--] \textbf{Proc assign},
\href{http://pubs.amstat.org/doi/abs/10.1198/106186001317114938}{http://pubs.amstat.org/doi/abs/10.1198/}
\href{http://pubs.amstat.org/doi/abs/10.1198/106186001317114938}{106186001317114938}
\item[]Can be used to perform optimal matching.

\item[--] \textbf{$\bolds{1\dvtx 1}$ Mahalanobis matching within propensity score calipers},
\href{http://www.lexjansen.com/pharmasug/2006/publichealthresearch/pr05.pdf}{www.lexjansen.com/pharmasug/2006/}\break
\href{http://www.lexjansen.com/pharmasug/2006/publichealthresearch/pr05.pdf}{publichealthresearch/pr05.pdf}
\item[]Feng, W. W., Jun, Y. and Xu, R. (2005). A method/\break macro based on
propensity score and Mahalanobis distance to reduce bias in treatment
comparison in observational study.

\item[--] \textbf{vmatch macro},
\href{http://mayoresearch.mayo.edu/mayo/research/biostat/upload/vmatch.sas}{http://mayoresearch.mayo.edu/mayo/}
\href{http://mayoresearch.mayo.edu/mayo/research/biostat/upload/vmatch.sas}{research/biostat/upload/vmatch.sas}
\item[]Kosanke, J. and Bergstralh, E. (2004). Match cases to controls using
variable optimal matching. Variable ratio matching (optimal algorithm).
\item[--] \textbf{Weighting},
\href{http://www.lexjansen.com/wuss/2006/Analytics/ANL-Leslie.pdf}{http://www.lexjansen.com/wuss/2006/}
\href{http://www.lexjansen.com/wuss/2006/Analytics/ANL-Leslie.pdf}{Analytics/ANL-Leslie.pdf}
\item[]Leslie, S. and Thiebaud, P. (2006). Using propensity scores to adjust
for treatment selection bias.
\end{itemize}
\end{longlist}

\section*{Acknowledgments}
Supported in part by Award K25MH083846 from the National Institute of
Mental Health. The content is solely the responsibility of the author
and does not necessarily
represent the official views of the National Institute of Mental Health
or the National Institutes of
Health.
The work for this paper was partially done
while the author was a graduate student at Harvard University,
Department of Statistics, and a Researcher at Mathematica Policy
Research. The author particularly thanks Jennifer Hill, Daniel Ho,
Kosuke Imai, Gary King and Donald Rubin, as well as the reviewers and
Editor, for helpful comments.

\vspace*{-2pt}
\end{document}